# What is the role of human decisions in a world of artificial intelligence: an economic evaluation of human-AI collaboration in diabetic retinopathy screening

**Authors:** Yueye Wang[1], Wenyi Hu[2, 3], Keyao Zhou[4], Chi Liu[5], Jian Zhang[6], Zhuoting Zhu[2, 3], Sanil Joseph[2, 3], Qiuxia Yin [6], Lixia Luo[6], Xiaotong Han [6]*, Mingguang He[1, 7, 8]*, Lei Zhang [9-12]*

**Affiliations:**

[1]School of Optometry, The Hong Kong Polytechnic University, Kowloon, Hong Kong.

[2]Centre for Eye Research Australia, Royal Victorian Eye and Ear Hospital, East Melbourne, Australia.

[3]Department of Surgery (Ophthalmology), The University of Melbourne, Melbourne, Australia.

[4]Department of Neurosurgery, Huashan Hospital, Fudan University, Shanghai, China.

[5]Faculty of Data Science, City University of Macau, Macao SAR, China.

[6]State Key Laboratory of Ophthalmology, Zhongshan Ophthalmic Center, Sun Yat-sen University, Guangdong Provincial Key Laboratory of Ophthalmology and Visual Science, Guangdong Provincial Clinical Research Center for Ocular Diseases, Guangzhou, China.

[7]Research Centre for SHARP Vision (RCSV), The Hong Kong Polytechnic University, Kowloon, Hong Kong.

[8]Centre for Eye and Vision Research (CEVR), 17W Hong Kong Science Park, Hong Kong.

[9]The Second Affiliated Hospital of Xi'an Jiaotong University, No.157 Xi Wu Road, Xi'an 710004, Shaanxi Province, PRC.

[10]China-Australia Joint Research Center for Infectious Diseases, School of Public Health, Xi'an Jiaotong University Health Science Center, Xi'an, Shaanxi, 710061, PR China

[11]Melbourne Sexual Health Centre, Alfred Health, Melbourne, VIC, Australia.

[12]Central Clinical School, Faculty of Medicine, Nursing and Health Sciences, Monash University, Melbourne, VIC, Australia.

*Corresponding author:

Lei Zhang. Email: lei.zhang1@monash.edu

Mingguang He. Email: mingguang.he@polyu.edu.hk

Xiaotong Han. Email: lh.201205@aliyun.com

**Abstract:** As Artificial intelligence has been increasingly integrated into the medical field, the role of humans may become vague. While numerous studies highlight artificial intelligence's potential, how humans and artificial intelligence collaborate to maximize the combined clinical benefits remains unexplored. In this work, we analyze 270 screening scenarios from a health-economic perspective in a national diabetic retinopathy screening program involving multiple human-artificial intelligence collaborative strategies and traditional manual screening. We find that a "copilot" decision-making approach, where decisions are made upon both-side agreement is the best strategy for human-artificial intelligence collaboration. It brings health benefits equivalent to US$ 4.64 million per 100,000 population. These findings demonstrate that even in settings where artificial intelligence is highly mature and efficient, human involvement remains essential to ensuring both health and economic benefits.

## Introduction

Artificial intelligence (AI) has shown promising capabilities in assisting humans in numerous healthcare settings, especially in detecting diseases from medical images at expert-level performance (*1*). A notable example is AI-assisted diabetic retinopathy (DR) screening, which has been successfully implemented in primary healthcare settings in the last decade (*2, 3*). DR is a major complication of diabetes and a leading cause to irreversible visual impairment. Among the more than 500 million adults living with diabetes worldwide (*4*), DR affected approximately 103 million people in 2020 and this number is projected to increase to 160 million by 2045 (*5, 6*). Many advanced AI tools for DR screening have shown strong diagnostic performances and have cleared regulatory approval into medical practice (*2, 3, 7-12*). Further, the health and economic impacts of these tools have been evaluated in various countries, demonstrating the strong cost-effectiveness of using AI (*13-22*).

While numerous studies have provided exciting advancements in the applications of AI, the role of human expertise in these settings remains essential. Although AI has demonstrated cost-effectiveness and efficacy in disease screening, it is still important to explore how human experts collaborate with AI to enhance its capabilities and maximise health impacts on its target population. AI excels in processing large volumes of data quickly and efficiently, particularly in repetitive tasks. However, human clinicians bring invaluable clinical experience that is crucial for handling complex and atypical cases. Together, AI and human expertise each offer unique strengths, making their collaboration essential for real-world clinical practices (*23, 24*).

Most previous DR screening programs have relied on AI models making independent decisions, with only a few models exploring a collaborative approach with human graders (*25, 26*). Further, in these models, very limited exploration has been done on the potential benefits of human-AI collaboration beyond AI's standalone performance. Only one study in Singapore found that combining AI with human screening was more cost-effective than using AI or humans alone, but this was a short-term cost-minimization analysis and did not assess long-running health outcomes of the strategy (*19*). To date, no research has discussed how to best structure human-AI collaboration to maximize the health-economic outcomes in DR screening. Lots of collaborative strategies can be explored. For example, humans need to confirm AI's results, or decisions can only be made upon human-AI consensus. Identifying the optimal strategy for human-AI collaboration is critical for maximizing the combined health-economic benefit and informing the best practice of AI-assisted interventions.

To fill this gap, we focused on China, the country with the largest number of diabetes cases worldwide and a burgeoning landscape of AI development (*27*). China presents a unique opportunity to evaluate the cost-effectiveness of various DR screening strategies in light of human-AI collaboration. Our study aims to assess the most comprehensive collaborative strategies from a healthcare provider's perspective, compared to the current practice. We employed a hybrid decision tree/Markov model to simulate the natural DR progression and explored diverse screening scenarios by varying screening intervals and target age groups. This study will provide key insights to optimize human-AI collaboration in real-world clinical practice.

# Results
## Performance of human-AI collaboration screenings
### *Performance of various screening strategies*

We compared the performance of manual screening and the eight AI-based screening strategies. 'AI triage before manual' (strategy 8, AI+M·M+M2, M represents primary human grader, M2 represents secondary human grader) achieved the highest overall accuracy of 99.49%, with 94.85% sensitivity and 99.86% specificity (Fig. 1). In contrast, the lowest accuracy was observed in 'AI' screening (strategy 1, accuracy: 81.89%; sensitivity: 96.15%; specificity: 80.74%). For sensitivity, 'Copilot human-AI' (strategy 6, AI·M+M2) achieved the highest at 99.27%, indicating it was the most capable at detecting positive cases. For specificity, 'sequential human review' (strategy 7, AI+M+M2) achieved the highest at 99.99%, reflecting its near-perfect ability to identify negative cases.

**Fig. 1. Multiple screening strategies for VTDR detection.**

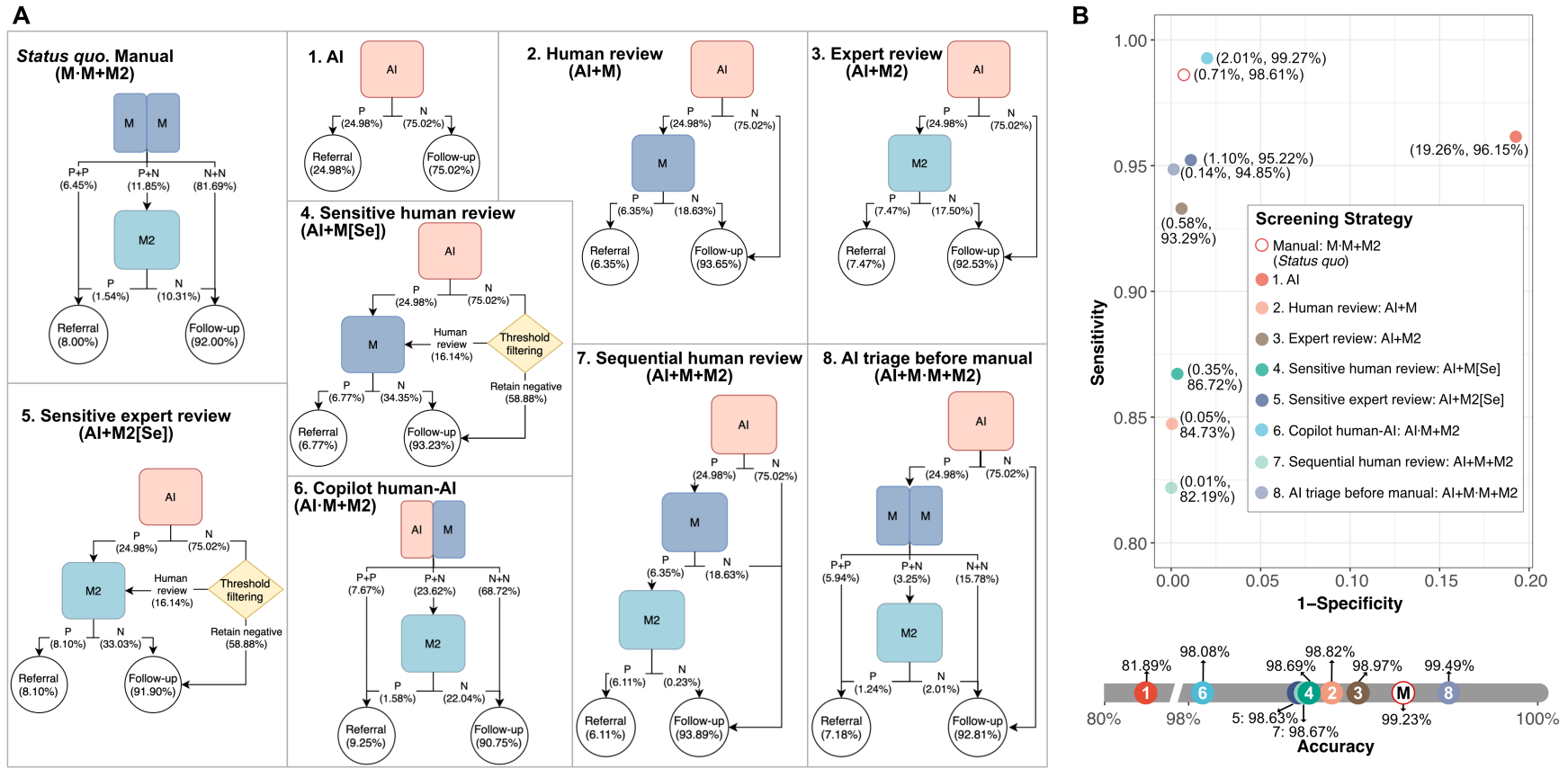

Notes: AI = artificial intelligence, M = primary grader, M2 = secondary grader, P = positive, N = negative. **A**. Structure of screening strategies. Positive pathway includes cases graded as either positive or ungradable. Patients' proportion in each pathway were listed according to each strategy's performance in the Lifeline Express Program. Names of screening strategies indicate their structure. In the name, the mark "·" represents both sides evaluation independently, decision will be made upon agreement. The mark "+" represents a superior level checking for positive or cases with disagreement. For example, in the status quo-"Manual (M·M+M2)" screening, two primary human graders graded all images independently; images with disagreement were sent to a secondary grader for review, whose

decision was considered as final. In strategy 2-"Human review (AI+M)" screening, a primary human grader checked all AI-positive cases. During threshold filtering in strategy 4-"Sensitive human review (AI+M[Se])" and 5-"Sensitive expert review (AI+M2[Se])", an AI score was computed as the probability of the image belonging to a particular category estimated by AI (e.g., the AI score for an R0 image is its probability of being classified as R0 by the AI). To maximize overall sensitivity, we determined that a negative image needs to have its AI score higher than 40% of all negative scores and have neither of its R2-R4 probabilities higher than 0.1. Otherwise, the image would be sent for human review. It can be recognized that the overall sensitivity of this strategy became higher after the threshold filtering process. **B**. Screening performance of the nine AI and manual screening strategies. Sensitivity and (1-specificity) were indicated for each strategy in a dot plot, while accuracies were plotted horizontally.

*Effectiveness of various screening strategies*

In the 100,000 population we modelled, manual screening (*status quo*) in different target groups with various screening frequencies would cost US$ 1666-1912 million and lead to 1,327,484-1,397,746 quality-adjusted life-years(QALYs) over lifetime. Compared to the status quo, 'sequential human review' (strategy 7: AI+M+M2) cost the most (extra US$ 7.39 [-2.05~18.21] million) while 'sensitive expert review' (strategy 5, AI+M2[Se]) screening is the most cost-reducing one (extra US$ -0.33 [-4.89~2.04] million saving). Regarding health effectiveness, only 'copilot human-AI' (strategy 6, AI·M+M2) screening gained extra 22-227 QALYs and achieved 67 to 658 additional years without blindness than status quo, with extra costs of US$ -0.47-0.90 million; while other AI-based screening led to less QALYs.

*Cost-effectiveness of various screening strategies*

Of all the 270 screening scenarios, the most cost-effective scenario was annual 'copilot human-AI' screening in the 20-79 age group, yielding an incremental cost-effectiveness ratio (ICER) of $6194/QALY gained and $2116/year per blindness year averted compared to the *status quo* (Fig. 2). The finding is regardless of using 1-time or 3-time per capita gross domestic product (GDP) as the willingness-to-pay (WTP) threshold. Broader comparisons were performed across screening frequencies in different target age groups, and indicated the robust cost-effectiveness of 'copilot human-AI' screening (fig. S1).

**Fig. 2. Cost-effectiveness frontier for annual screening scenarios in 20-79 target age group in 100,000 population with diabetes over lifetime.**

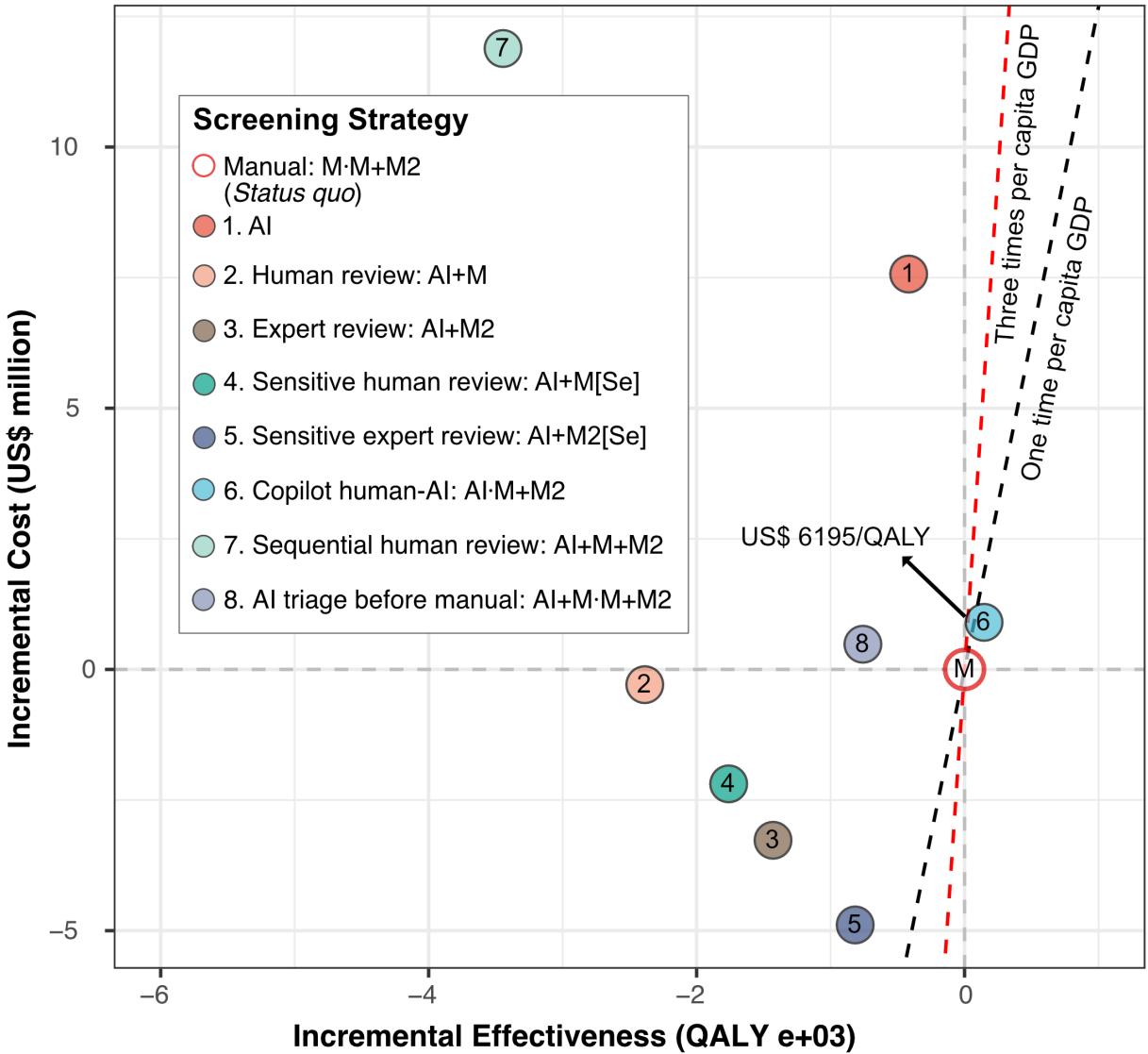

Notes: QALY = quality-adjusted life-years. GDP=gross domestic product. Costs (US$) and effectiveness (QALYs) of different screening strategies were compared with the status quo (manual screening). The strategies on the upper left of the frontier are dominated by the strategies on the lower right of them. "Copilot human-AI (AI·M+M2)" screening is very cost-effective, and the only cost-effective strategy compared to the status quo.

Higher screening frequency and larger target population age lead to greater cost-effectiveness. Annual screening in the 20-79 age group required lower costs and showed higher health effectiveness than other frequencies and other age groups (fig. S2). Therefore, we used this screening setting for effectiveness comparison between different human-AI collaborative strategies with the *status quo*. As shown in Table 1, using the most cost-effective strategy - 'copilot human-AI screening', to replace manual screening, would require an additional investment cost of $0.9 million but gain additional 146 QALYs and 426 years without blindness in the entire population. The health benefit from this strategy was equivalent to US$ 4.64 million. Implementing this strategy could prevent 33 blindness cases, result in 115 additional vision-threatening diabetic retinopathy (VTDR) cases being diagnosed, and ensure that 80 of these cases receive treatment.

**Table 1. Comparison with the status quo on cost-effectiveness for AI-based DR screening strategies for annual screening in 20-79 age group.**

| Cost/Effectiveness of annual screening scenarios | Status quo<br><br>Manual<br><br>M·M+M2 | Incre. cost/effectiveness comparison to the status quo | | | | | | | |
|---|---|---|---|---|---|---|---|---|---|
| | | 1<br><br><br><br>AI | 2<br>Human review<br><br>AI+M | 3<br>Expert review<br><br>AI+M2 | 4<br>Sensitive human review<br>AI+M[Se] | 5<br>Sensitive expert review<br>AI+M2[Se] | 6<br>Copilot human-AI<br>AI·M+M2 | 7<br>Sequential human review<br>AI+M+M2 | 8<br>AI triage before manual<br>AI+M·M+M2 |
| **VTDR cases** | | | | | | | | | |
| Blindness cases | 43,463 | 94 | 543 | 324 | 400 | 185 | -33 | 787 | 172 |
| Detected VTDR cases | 74,747 | -329 | -1893 | -1131 | -1394 | -645 | 115 | -2741 | -599 |
| Treated VTDR cases | 52,323 | -230 | -1325 | -792 | -976 | -451 | 80 | -1919 | -419 |
| **Costs (US$ million)** | | | | | | | | | |
| Screening cost | 32.80 | -14.80 | -10.28 | -9.13 | -10.25 | -8.75 | -0.27 | -1.14 | -1.76 |
| Referral cost | 16.59 | 20.60 | -0.13 | -0.21 | 0.59 | 0.40 | 1.79 | -1.56 | -0.98 |
| Treatment cost | 451.12 | -2.39 | -13.68 | -8.19 | -10.08 | -4.67 | 0.83 | -19.76 | -4.34 |
| Blindness cost | 1165.47 | 4.16 | 23.80 | 14.26 | 17.55 | 8.14 | -1.45 | 34.34 | 7.57 |
| Total cost | 1665.98 | 7.57 | -0.29 | -3.27 | -2.19 | -4.89 | 0.90 | 11.89 | 0.48 |
| **Effectiveness** | | | | | | | | | |
| QALYs | 1,397,746 | -417 | -2387 | -1430 | -1760 | -816 | 146 | -3445 | -759 |
| Years without blindness | 1,418,316 | -1221 | -6985 | -4184 | -5152 | -2390 | 426 | -10,079 | -2221 |
| **Cost-effectiveness evaluation** | | | | | | | | | |
| ICER (Incre.l costs/Incre. QALYs) | - | Dominated | 122 | 2287 | 1244 | 5989 | 6194 | Dominated | Dominated |
| Cost/blindness year adverted (Incre. | - | Dominated | 42 | 781 | 425 | 2046 | 2116 | Dominated | Dominated |

| costs/Incre. years without blindness) | | | | | | | | |
|---|---|---|---|---|---|---|---|---|
| NMB (US$ million) | - | -23.44 | -90.53 | -51.13 | -64.80 | -26.18 | 4.64 | -142.97 | -29.35 |

Notes: # Cost saving scenarios compared to manual screening. Incre. = Incremental, DR = diabetic retinopathy, AI = artificial intelligence, VTDR = vision-threatening referable DR, QALY = quality-adjusted life-years, ICER = incremental cost-effectiveness ratio, NMB = net monetary benefit.

Consistent trends were observed across all age groups. As the initial screening age increased, AI-based strategies generally became more cost-effective compared to the *status quo*. However, it was only the 'copilot human-AI screening' showed consistently higher effectiveness than the *status quo* across all age groups or screening frequencies.

**Sensitivity analysis**

Probabilistic sensitivity analysis revealed that for annual screening in the 20-79 age group, 'copilot human-AI' screening was the optimal strategy at 3 times per capita GDP (46.88% probability of being the most cost-effective strategy); yet if the WTP level was limited at 1 time per-capita GDP, 'sensitive expert review' screening became the optimal (40.33% probability of being the most cost-effective strategy) (Fig. 3). Compared with the *status quo*, only 'copilot human-AI' screening showed cost-effectiveness in most of the 10,000 simulations at the WTP level of either 1-time or 3-time per-capita GDP, while other strategies were dominated by manual screening in over half of the simulations (Fig. 4).

**Fig. 3. Cost-effectiveness acceptability curve.**

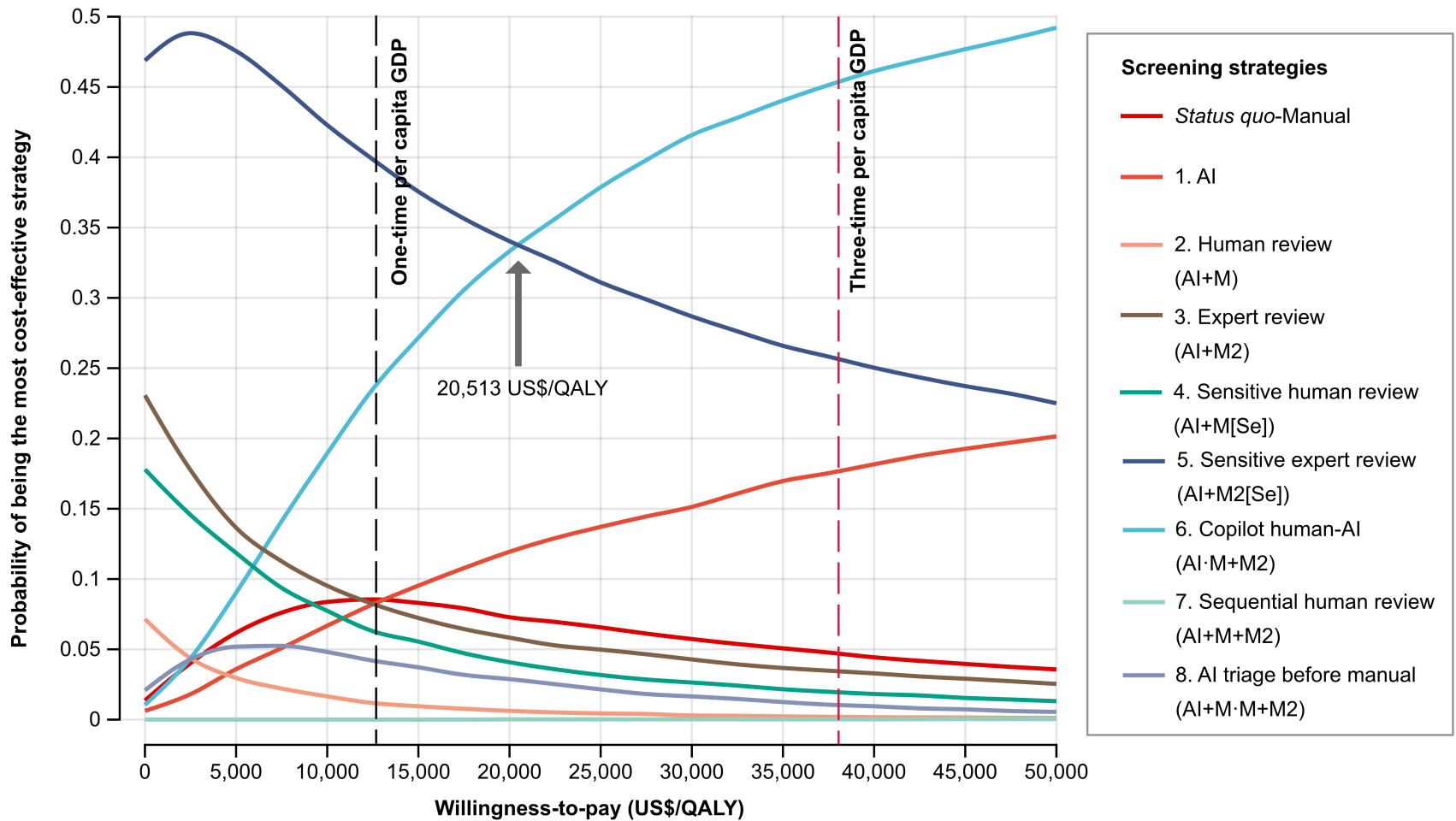

Notes: QALYs=quality-adjusted life-years. GDP=gross domestic product. Annual screening strategies for 20-79 age group were compared. When the WTP is lower than 1-time per-capita GDP, "Sensitive expert review (AI+M2[Se])" has the highest probability of

being cost-effective; while at WTP higher than 3-time per-capita GDP, "Copilot human-AI (AI·M+M2)" become the most cost-effective choice. The switching point between these two strategies is shown at WTP of US$ 20,513 per QALY gained.

**Fig. 4. Probabilistic sensitivity analysis scatterplot.**

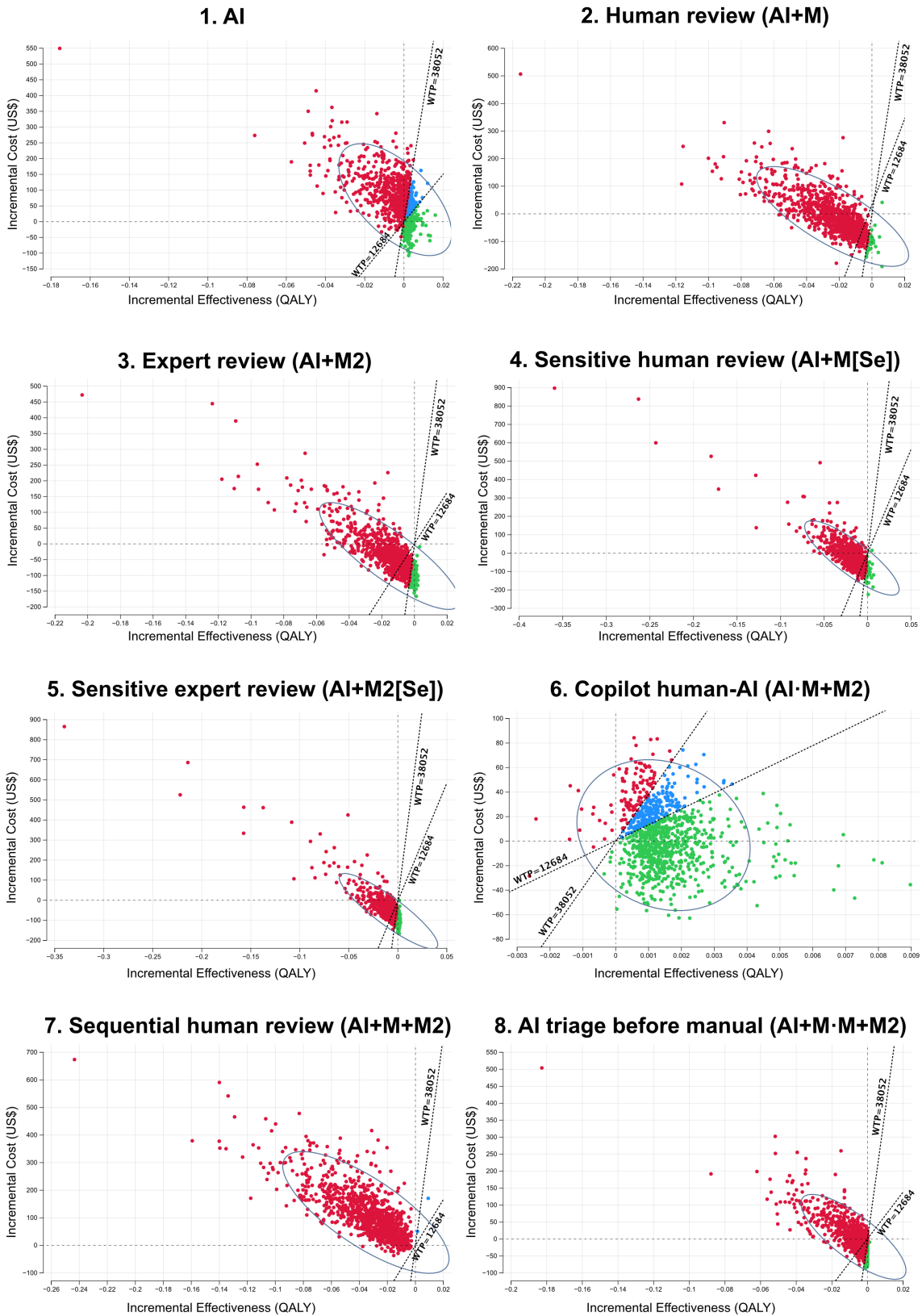

Notes: QALYs=quality-adjusted life-years, WTP = willingness-to-pay. Each of the eight human-AI collaborative screening strategies was compared with the status quo. Red dots represent simulations where the status quo (manual screening) dominates the AI-based strategy at a WTP threshold of 3 times per capita GDP. Blue dots indicate simulations where the AI-based strategy is cost-effective compared to the status quo at WTP levels between 1-3 times per capita GDP. Green dots represent simulations where the AI-based strategy is cost-effective at WTP levels below 1-time per capita GDP. Among all the screening strategies, only the "Copilot human-AI (AI·M+M2)" approach (panel 6) dominated the status quo in more than half of the simulations.

The univariate sensitivity analysis revealed that varying major sensitive parameters did not significantly change the cost-effectiveness ranking of different screening strategies. We compared the most cost-effective strategy (ie, 'copilot human-AI') to the other strategies under varying model parameters. The most influencing factors included the diagnostic performance of human graders and the AI system, as well as the cost for blindness care in follow-up years (fig. S3). 'Copilot human-AI screening' remained to be the most cost-effective strategy despite varying model parameters. The only case when 'sensitive expert review' screening became more cost-effective than 'copilot human-AI' screening was when the sensitivity of the secondary grader exceeded 97.9%.

Across all screening timeframes from 5 to 30 years for annual screening in the 20-79 age group, still, only the 'copilot human-AI' screening strategy demonstrated greater effectiveness compared to the *status quo* (fig. S4). However, this strategy was deemed cost-effective (ICER lower than WTP of 3-time per capita GDP) only when the screening timeframe exceeded 10 years. This strategy became highly cost-effective (ICER below the WTP threshold of 1-time the per capita GDP) when the timeframe extended beyond 20 years.

Furthermore, we evaluated cost-effectiveness for annual screening in the 20-79 age group from a health-care provider's perspective with no blindness-related societal costs involved. Compared to the *status quo*, sole AI and 'copilot human-AI' screening incurred higher costs, while other human-AI collaboration strategies were less expensive. Despite the higher cost, only the 'copilot human-AI' strategy provided greater effectiveness than the *status quo* at an ICER of US$ 16,170/QALY, making it remains to be the optimal cost-effectiveness strategy.

## Discussion

After analyzing 270 screening scenarios in a national DR screening program in China, we found that the strategy of annual 'copilot human-AI' (AI·M+M2) screening in the 20-79 age group was the most cost-effective. In this 'copilot human-AI' screening, AI and a primary human grader make independent decisions in parallel, afterwards a secondary grader reviews and resolves disagreements. This remained robust regardless of whether the WTP threshold was set at 1-time or 3-time per-capita GDP.

While AI models exhibit remarkable diagnostic accuracy, offering clear advantages in speed and cost, our findings underscore that relying solely on AI is not enough. In alignment with our findings, most investigations proposed that exclusive AI screening offered limited effectiveness gain compared to traditional human-based manual screening (*14, 17, 18*). Further, real-world AI may successfully grade only 87% of fundus images without pupil dilation (*28*). The need to refer ungradable cases and the inevitable false positive rate can result in unnecessary referrals and waste time and medical resources. Moreover, legal and ethical considerations surrounding the sole use of AI in medical diagnosis need to be addressed (*29, 30*). Combining AI efficiency and human expertise is essential. Our study, to the best of our knowledge, included all feasible human-AI collaborative strategies from previous research and real-world practice, providing the most comprehensive comparison analysis to date.

One study in the UK reported that in a sequential screening structure involving three human graders, implementing AI as either a replacement (similar to 'sequential human review' [AI+M+M2] screening in our study) or a filter before the primary grader can save costs but provide limited effectiveness (*21*). A cost-minimization study in Singapore recommended semi-AI screening (similar to 'human review' [AI+M] screening in our study) as this strategy was the least expensive compared to manual and AI screening (*19*). Consistent with the Singapore study, Liu et al. derived comparable findings in China, highlighting that semi-AI screening incurred lower costs while demonstrating higher effectiveness than the manual approach (*13*). Nevertheless, it is important to acknowledge that Liu's study targeted a general population, which differs from our specific focus on individuals with diabetes.

Determining the most cost-effective approach to implement AI is necessary to stay in step with advancing technology. Our findings recommend that the most cost-effective strategy, the 'copilot human-AI' screening strategy, should involve independent decision-making of AI and humans. However, this strategy has received minimum discussion before. AI algorithms are typically optimized for high sensitivity to maximize DR detection ability during the development stage (*31, 32*). Conversely, human graders may vary in their sensitivity in real-world grading tasks (often below 90%), but can maintain high specificities over 95% (*3, 33*). Therefore, the 'copilot human-AI' screening strategy effectively combines the advantages of both AI and human graders, maximizing the detection of positive cases and enabling timely intervention. The findings of this study inspire the real-world application of AI-based DR screening, with the potential to inform public health policy as well as encourage us to reconsider the role of AI in healthcare. The future of medical AI should prioritize a human-centric approach, focusing on in-depth collaboration between AI and healthcare providers.

Consistent with the current DR screening recommendations (*34*), our findings indicate that annual screening in the 20-79 age group is more cost-effective than targeting older populations or extending the screening interval. We also found that AI-based strategies offer clear cost-saving benefits when the initial screening age increases, as targeting the older population shortens the overall screening duration. Some previous studies demonstrated that sole AI screening can be more cost-effective than manual screening over a short period, such as five years (*15, 18, 20*). Combined with our findings, it can be hypothesized that AI may show cost-saving benefits and high effectiveness in the short term, yet long-term screening would require more human-AI collaboration to ensure sufficient health gain.

The primary strength of our study is the inclusion of the most comprehensive human-AI collaborative strategies for DR screening to date. The reliability and robustness of our results were also enhanced by using a national real-world data. However, it is essential to view the findings of our study within the context of its limitations. Firstly, our cost calculations for AI deployment only encompassed expenses related to development and IT maintenance as it is an academic version. It is worth noting that the cost of deploying commercial AI devices may vary widely, and are typically higher than that of the academic versions (*35*). This is a common limitation of current cost-effectiveness studies. Given the enormous investment required for deploying commercial AI systems in DR screening, per-patient benefits may be diminished. Secondly, our Markov model did not include diabetic macular edema and impaired visual acuity as referable health states. This omission arose from the limitations of real-world data, allowing us to simulate only a simplified DR progression. Future studies involving more detailed DR states are warranted. Thirdly, this study was conducted in China, a low- and middle-income country. External validation may be needed before extending these results to low- or high-income countries, or regions with different DR prevalence.

In conclusion, our results indicate that annual DR screening using the 'copilot human-AI' screening strategy in the 20-79 age group is the most cost-effective strategy in the current context of China. We believe our study findings provide important data reference and evidence for optimizing human-AI collaboration in developing country settings. Future advancements in AI hold promise in improving performance and cost-effectiveness of screening strategies, which could shift the most cost-effective option, and also may bring solely AI-based DR screening down to earth.

## Materials and Methods Summary

This study used data from Lifeline Express, a national DR screening program in China involving 251,535 participants from over 200 hospitals across the country *(26)*. A hybrid decision tree/Markov model was constructed to simulate DR screening scenarios. The model consisted of a hypothetical cohort of 100,000 diabetic individuals aged 18-79 years, with a yearly time-step over a life expectancy of 80 years.

Individuals are classified into five distinct health states, consistent with the National Health Service (NHS) DR grading guideline *(36)*: non-vision-threatening DR (non-VTDR), VTDR, treated DR, blindness, and death. VTDR encompasses severe non-proliferative DR and proliferative DR *(37)*.

Our model investigated a combination of nine distinctive screening strategies under six screening frequencies (one-off, annual, once every 2 years, once every 3 years, once every 4 years, and once every 5 years) and five screening age groups (20–79, 30–79, 40–79, 50–79, and 60–79 years). In total, our model investigated 270 screening scenarios (9 strategies × 6 frequencies × 5 age groups).

Based on the current practice, we set manual screening according to the NHS grading workflow as the *status quo*. In the *status quo* (manual: M·M+M2), two independent primary graders (M·M) review all images, followed by a secondary grader (M2) resolving disagreements as a final reviewer. '·' indicates independent decisions in parallel and '+' indicates sequential decision by reviewing a subset of cases from the previous step.

Eight human-AI collaborative strategies were selected according to general DR screening practice and existing literature *(13, 19, 21)* as follows (Fig. 1):

Strategy 1 (AI) involves AI making decisions independently.

Strategy 2 (Human review: AI+M) and strategy 3 (Expert review: AI+M2) both combine AI and a human grader (primary grader M or secondary grader M2) to review AI-positive cases.

Strategy 4 (Sensitive human review: AI+M[Se]) and strategy 5 (Sensitive expert review: AI+M2[Se]) introduce a threshold filtering process. Threshold filtering aims to improve screening sensitivity by ensuring that only images with high AI confidence as negative are classified as negative, while uncertain and positive cases are flagged for human review (primary grader M or secondary grader M2).

Strategy 6 (Copilot human-AI: AI·M+M2) uses AI to replace one primary grader (M) in manual screening. In this approach, AI and a primary grader (M) make independent decisions in parallel, and a second grader (M2) resolves disagreement.

Strategy 7 (Sequential human review: AI+M+M2) introduces a sequential review where AI, a primary grader (M), and a second grader (M2) sequentially review positive cases from the previous step.

Strategy 8 (AI triage before manual: AI+M·M+M2) adds an AI triage before manual screening, while manual screening will only be performed for AI-deemed positive cases.

We have also tried to list exhaustive AI-based screening strategies and finally confirmed other human-AI collaboration strategies are not practical or inferior to the eight ones we selected above in reality.

The incremental costs and incremental QALYs for each screening scenario were calculated compared to the *status quo* (manual screening) at the same screening frequency and target age group. ICER was defined as the incremental cost per QALY gained, cost per year of blindness averted was also calculated:

$$ICERs = \frac{incremental\ cost}{QALYs\ gained} \quad (1)$$

$$cost\ per\ blindness\ year\ averted = \frac{incremental\ cost}{years\ of\ blindness\ averted} \quad (2)$$

To obtain the most cost-effective strategy, we identified the cost-effectiveness frontier by comparing each screening scenario to the next best scenario (lower-cost non-dominated scenario). In accordance with World Health Organization (WHO) guidelines, our benchmarks for WTP thresholds were determined at 3-time per capita GDP as US$ 38,052, based on the 2023 overall per capita national GDP in China (US$12,684) *(38)*. Scenarios with an ICER <1, 1−3, and > 3 times the per capita GDP were defined as very cost-effective, cost-effective, and not cost-effective, respectively.

We performed one-way sensitivity analysis to examine the influence of model parameters on the outcome. Probabilistic sensitivity analysis represents the probability of each screening scenario being cost-effective compared with all others. The analysis was based on 10,000 Monte-Carlo simulations, assuming a variety of probability distributions governing model parameters. To assess the differences between short-term and long-term screening programs, we varied the timeframes and calculate cost-effectiveness of each AI-based screening strategy against the *status quo* for annual screening in 20-79 age group. The timeframes under consideration ranged from 5 to 30 years. Additionally, we assessed cost-effectiveness from a healthcare provider's perspective by excluding costs associated with blindness care during the initial and follow-up years.

**Fig. S1.**

Cost-effectiveness frontiers across screening frequencies in different target age group in 100,000 population with diabetes over lifetime.

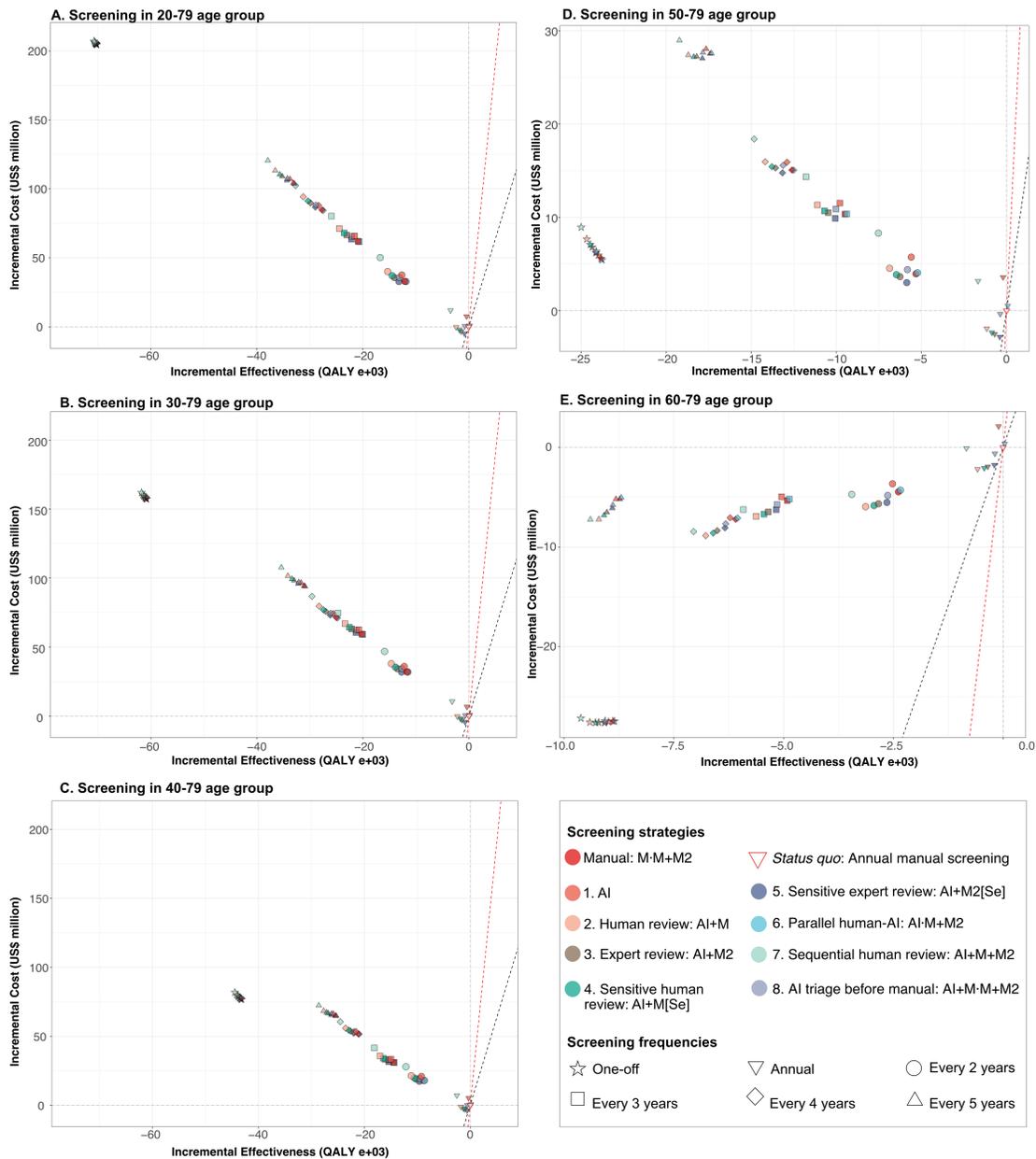

Notes: QALY = quality-adjusted life-years. GDP=gross domestic product. Costs (US$ million) and effectiveness (QALYs e03) of eight AI-based screening strategies and manual screening across different screening frequencies were compared with annual manual screening as the *status quo*. These comparisons were performed in 5 target age groups (20–79, 30–79, 40–79, 50–79, and 60–79 years). The strategies on the upper left of the frontier are dominated by the strategies on the lower right of them. Black dashed line indicates WTP at one time per capita GDP, while red dashed line indicates WTP at three-time per capita GDP. In all age groups, "Copilot human-

AI (AI·M+M2)" screening was found very cost-effective, and the only cost-effective strategy compared to the *status quo*.

**Fig. S2.**

Comparison in cost and effectiveness across screening frequencies and target age groups.

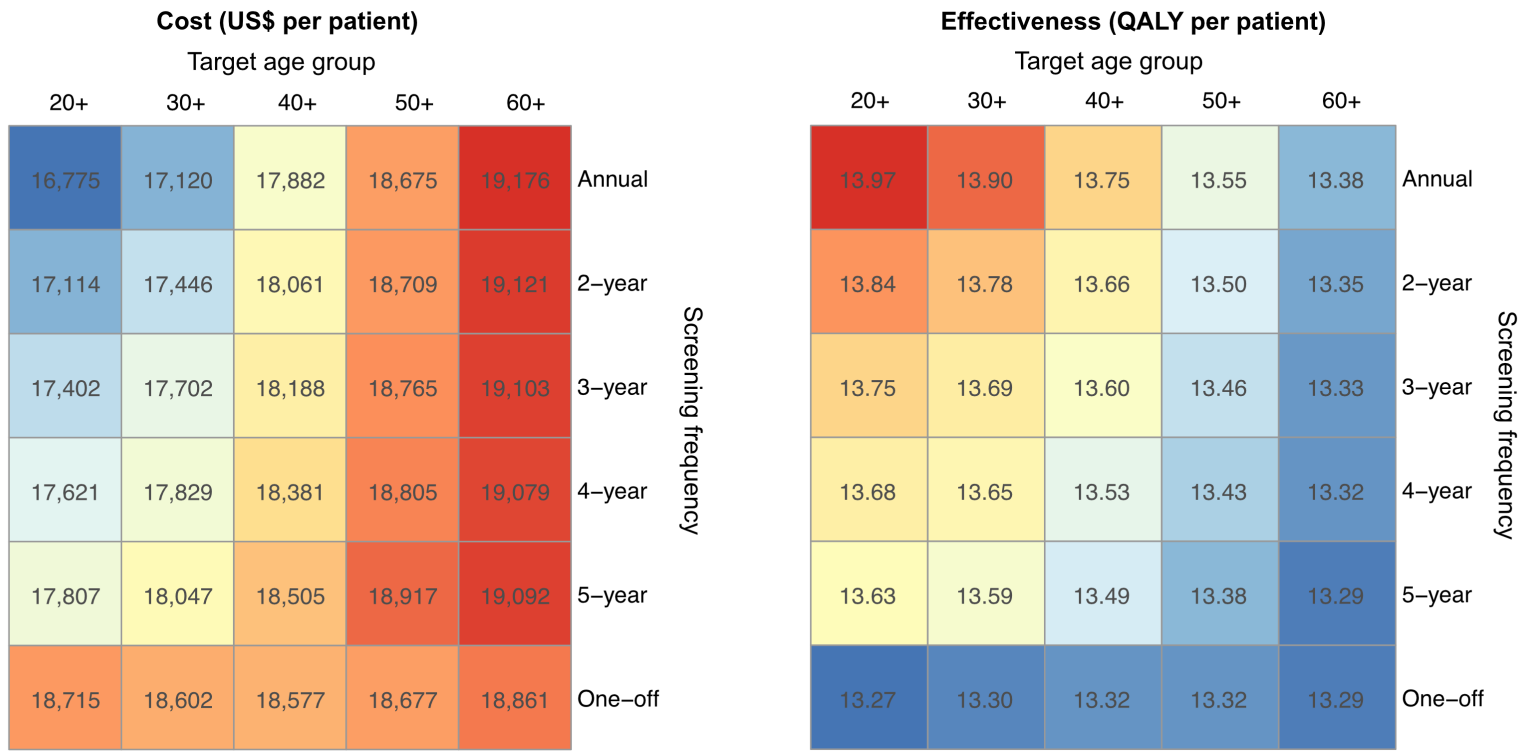

Notes: Costs and effectiveness were averaged for different screening strategies.

**Fig. S3.**

One-way Tornado diagram comparing the most cost-effective strategy with others in the reference setting.

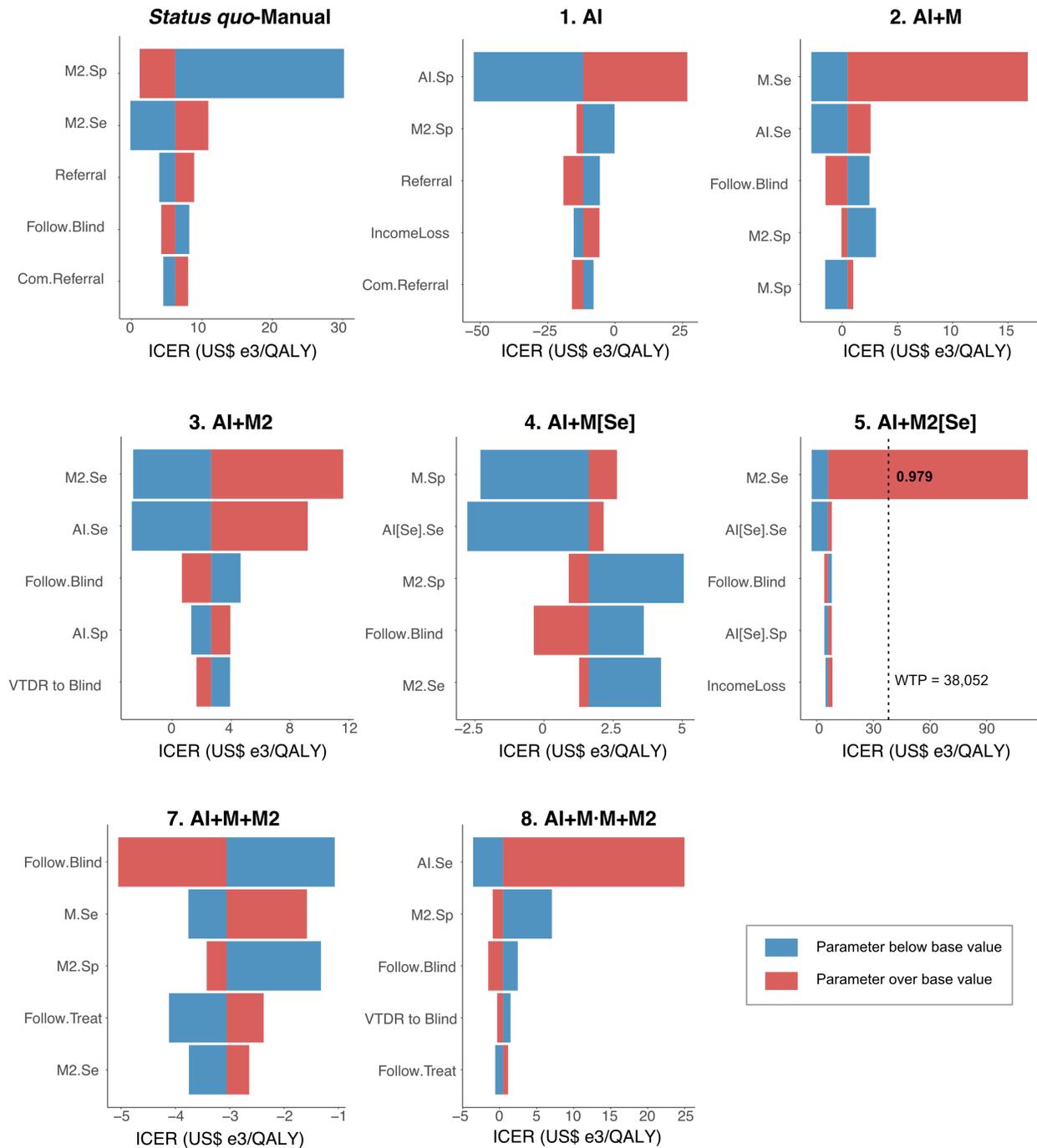

Notes: AI = artificial intelligence, VTDR = vision-threatening diabetic retinopathy. Sp. = specificity, Se. = sensitivity, M = primary grader, M2 = secondary grader, Follow. = costs during follow-up, Treat = Treated VTDR, AI[Se] = AI system after threshold adjustment, WTP =

willingness-to-pay, ICER = incremental cost-effectiveness ratio, QALY = quality-adjusted life-year. GDP = gross domestic product.

The reference group was determined at the 20-79 age group, screening annually. ICERs are calculated as QALYs divided by costs (US$/QALY). One-way sensitivity results of the top five parameters that had the largest effect on ICER are shown for each screening strategy, compared with "AI·M+M2" screening at an interval of annual screening in 20-79 age group. The scenario was defined as cost-effective if the ICER was less than three times per-capita GDP (US$ 38,052). The "AI·M+M2" screening every year is maintained to be the most cost-effective strategy despite varying model parameters. The only case when "AI+M2[Se]" screening becomes more cost-effective than " AI·M+M2" screening is when the sensitivity of the secondary grader exceeds 0.979.

**Fig. S4.**

Cost-effectiveness of multiple screening strategies across different timeframe.

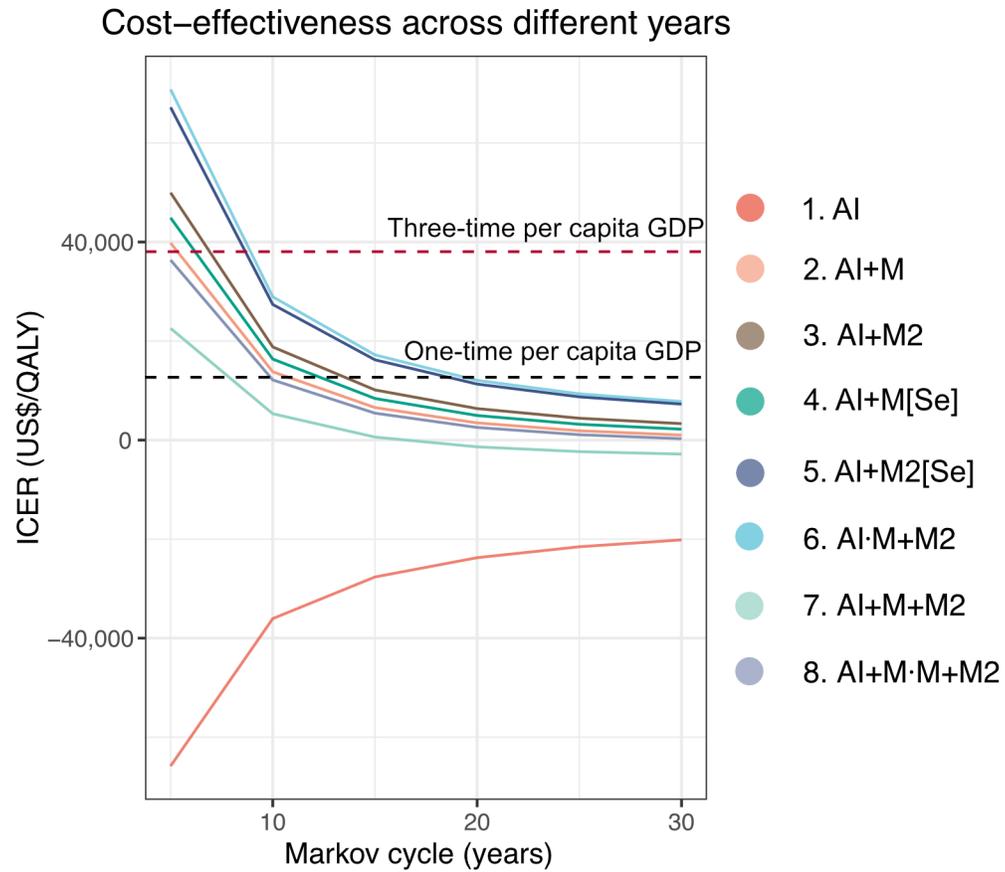

Notes: AI = artificial intelligence, M = primary grader, M2 = secondary grader, AI[Se] = AI system after threshold adjustment, WTP = willingness-to-pay, ICER = incremental cost-effectiveness ratio, QALY = quality-adjusted life-year.